# ProTranslator: zero-shot protein function prediction using textual description

Hanwen Xu[1], Sheng Wang[2*]
[1]Department of Automation, Tsinghua University, Beijing, China, email: xuhw20@mails.tsinghua.edu.cn
[2]Paul G. Allen School of Computer Science & Engineering, University of Washington, Seattle, WA
*Correspondence to swang@cs.washington.edu

**Abstract**
Accurately finding proteins and genes that have a certain function is the prerequisite for a broad range of biomedical applications. Despite the encouraging progress of existing computational approaches in protein function prediction, it remains challenging to annotate proteins to a novel function that is not collected in the Gene Ontology and does not have any annotated proteins. This limitation, a "side effect" from the widely-used multi-label classification problem setting of protein function prediction, hampers the progress of studying new pathways and biological processes, and further slows down research in various biomedical areas. Here, we tackle this problem by annotating proteins to a function only based on its textual description so that we don't need to know any associated proteins for this function. The key idea of our method ProTranslator is to redefine protein function prediction as a machine translation problem, which translates the description word sequence of a function to the amino acid sequence of a protein. We can then transfer annotations from functions that have similar textual description to annotate a novel function. We observed substantial improvement in annotating novel functions and sparsely annotated functions on CAFA3, SwissProt and GOA datasets. We further demonstrated how our method accurately predicted gene members for a given pathway in Reactome, KEGG and MSigDB only based on the pathway description. Finally, we showed how ProTranslator enabled us to generate the textual description instead of the function label for a set of proteins, providing a new scheme for protein function prediction. We envision ProTranslator will give rise to a protein function "search engine" that returns a list of proteins based on the free text queried by the user.

## 1 Introduction

Accurately identifying protein functions serves as the basis for studying a wide range of biomedical problems [1-4], such as cell cycle regulation [5], neuronal morphogenesis [6], signal transduction [7] and drug discovery[8,9]. Experimentally testing the functions of millions of proteins across tens of thousands of functions is impractical. As a result, many computational approaches have been proposed to predict protein functions according to protein domains [1,10], protein motifs [11,12], protein sequence [13-17], protein-protein interactions [18-20], protein text description [21], and protein structures [22,23]. These features have also been integrated to jointly perform the prediction [24-27]. The standard problem setting of protein function prediction is to form it as a multi-label classification problem, where the input is the feature vector of a protein and the output is a set of functions predefined as controlled vocabulary in the Gene Ontology (GO) [28]. This problem setting enables protein function prediction to easily incorporate new machine learning techniques in the feature extraction or the classification component, but inevitably restricts the predicted function to be within the set of controlled vocabulary. As a result, existing methods are not able to classify proteins into functions that are not in the GO and do not have any annotated proteins. clusDCA is able to classify proteins to the function that do not have any annotated proteins [18], but it still requires that function to be within the GO graph. This limitation substantially hinders the progress towards understanding new molecular functions and biological processes, further slowing down research in downstream applications.

We aim to develop an algorithm that enables us to classify proteins into any function that does not have any annotated proteins and is not in GO. The only information we need for that function is a textual description, which could be a few sentences describing this function. The key idea of our method is to embed descriptions of all GO functions into the same low-dimensional space, where similar functions are co-located. When we need to annotate a new function that is not in GO, we will project this new function in this low-dimensional space based on its textual description and then transfer annotations from other GO functions. To embed the textual description, we used large-scale language model PubMedBert [29], which is pre-trained on millions of scientific papers and obtained the state-of-art performance on specialized biomedicine tasks. We then embed proteins by integrating protein sequence, protein textual description and protein-protein interaction network. Finally, we learnt a linear transformation from protein embedding space to function embedding space according to known annotations.

We validated our method on CAFA3 [2], SwissProt [30], and GOA [28] datasets and observed substantial improvement on functions that do not have any annotations and functions that are sparsely annotated. We further demonstrated how our method could predict gene members of pathways in Reactome [31], KEGG [32], MsigDB [33] by only using the pathway description without seeing any specific gene belonging to it. Finally, we showed how to generate sentences that can best describe the function of a set of given proteins based on our method. We envision that our method enables us to build a "search engine" for function prediction, where users only need to provide a few keywords or sentences to describe the function that they want to annotate and then our system will return the associated proteins and genes for this function.

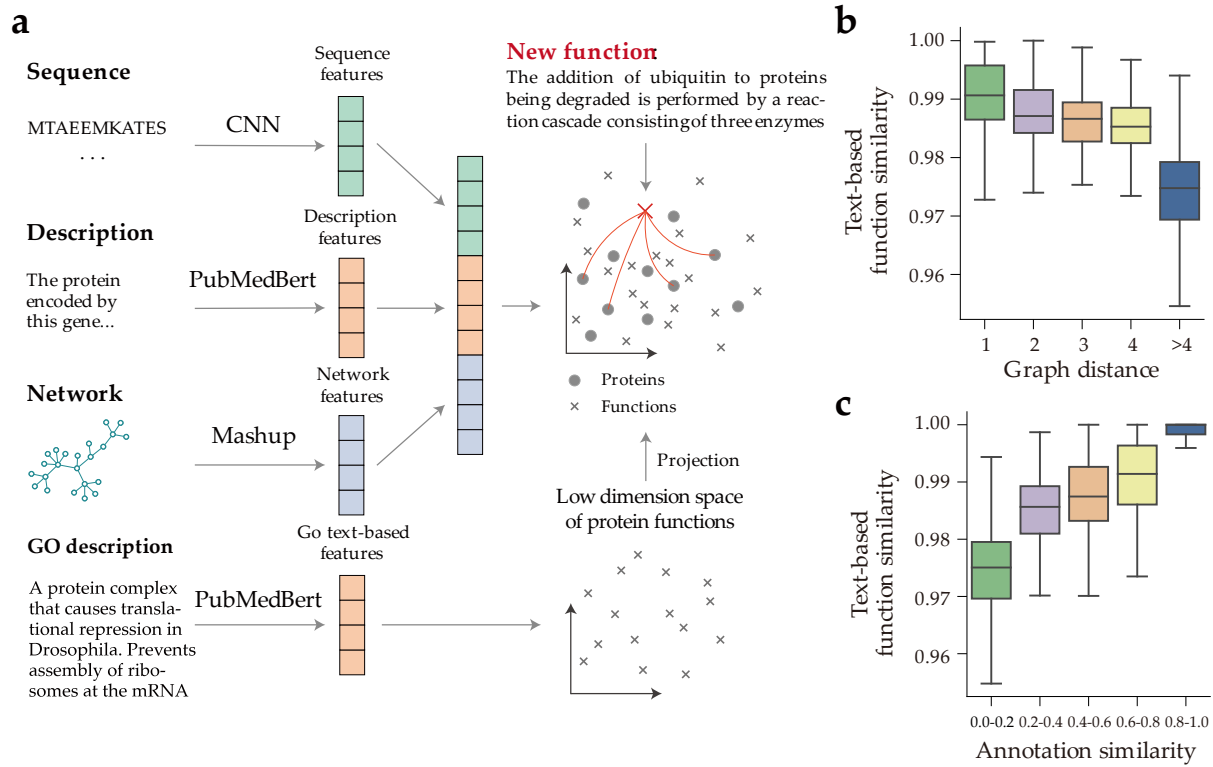

**Fig. 1. a**, Flowchart of ProTranslator. ProTranslator embeds proteins into the low-dimensional space by integrating protein sequence, description and network features. It then embeds GO terms according to the textual description. GO terms are then projected into the protein embedding space according to known annotations. To annotate a new function, ProTranslator will project the new function into this low-dimensional space according to its textual description and annotate it using nearby proteins. **b,c,** Box plots comparing text-based GO similarity for GO terms that have different distances on the graph (**b**) and annotation similarity (**c**).

## 2 Methods

### 2.1 Problem definition

Previous studies modelled the protein function prediction as a multi-label classification problem. They used features of proteins, denoted as $X$, to predict a subset $Y$ of predefined functions $Y_0 = \{y_1, y_2, \ldots, y_{z_0}\}$ where $z_0$ denotes the number of predefined functions. For each function $y_j$ in $Y_0$, it has annotated protein feature set $X_j$ collected for the model training. However, this modelling approach restricts the scope of protein prediction research to the function within the controlled vocabulary $Y_0$. To tackle this problem, we redefine this problem by considering the potential novel function set without any annotated protein collected. The novel function set is defined as $U_0 = \{u_1, u_2, \ldots, u_{z_1}\}$ and $z_1$ denotes the number of novel functions. There is also a new protein feature set $X^U{}_j$ for each $u_j$ and $X^U{}_j \notin \{X_1, \ldots, X_{z_0}\}$. This problem is defined as: with only protein features $\{X_1, \ldots, X_{z_0}\}$ and their annotations $\{Y_1, Y_2, \ldots, Y_{z_0}\}$ seen before, the prediction method should learn to classify a new protein feature set $X^U{}_j$ into a novel function $u_j$.

### 2.2 Embedding GO functions based on the textual description

We use the textual description to embed GO functions. They are collected from the "definition" field from the Gene Ontology. In order to embed a new function to the same low-dimensional space only based on textual description, we disregard other information, such as GO graph and protein annotations, when embedding GO functions. We obtain the vector representation for each GO term using PubMedBert, which is pre-trained on both the PubMed's abstracts and PubMedCentral's full-text articles [29]. The corpus used by PubMedBert is best aligned with our task in comparison to other pre-trained language models. To obtain a fixed-size feature vector for GO definitions with various numbers of words, We average the last hidden states' output on each dimension across all tokens (words or subwords) to obtain the low-dimensional text representations. The final representation vector for each GO term is $d_{Bert}$ dimensions.

### 2.3 Embedding proteins based on sequence, description and network

To embed proteins, we consider three widely-used features: sequence, description and network. We followed the state-of-the-art approach DeepGOPlus to extract the sequence features using convolution neural networks (CNN) [14]. Multiple 1-d convolution kernels with different sizes are used in the first layer and the size ranges from $R_D$ to $R_U$ with step $T$, which results in $e$ different sizes. The number of filters of each size is set to $d_0$. Then a max pooling layer is used to extract information across kernels.

$$FS_t = f_1([\omega_{1,t} * x^s{}_{1:t}, \omega_{1,t} * x^s{}_{2:1+t}, \ldots, \omega_{1,t} * x^s{}_{L-t+1:L}]), \quad (1)$$

where $\omega_{1,t}$ represents the 1-d convolution kernel in the first layer with the window size $t$ and $x^s \in \mathbb{R}^{n \times 21 \times L}$ represents the input sequence one-hot encodings. $n$ denotes the number of proteins. $f_1$ denotes the max pooling layer with the kernel size $L - t + 1$. We then concatenate different $FS_t \in \mathbb{R}^{n \times d_0}$ together as the sequence features $FS \in \mathbb{R}^{n \times d_{seq}}$.

$$FS = concat(FS_{t_1}, FS_{t_2}, \ldots FS_{t_e}), t_1 \leq t_2 \leq \ldots \leq t_e \leq L. \quad (2)$$

We embed the protein descriptions similarly to the process of embedding the textual description of GO functions. The description of each protein was obtained from GeneCards [34,35]. Each protein is then represented by a low-dimensional vector $FD \in \mathbb{R}^{d_{Bert}}$. The gene network data $FN \in \mathbb{R}^{d_{Mashup}}$ is provided by the pre-trained Mashup representations of each protein according to their topology in multiple protein-protein interaction networks [20]. Then we add one-depth fully connected layers to reshape each kind of feature ($FS$, $FD$ and $FN$) and set the output dimension to $h$. The fully connected layers are denoted as $Layer_{FCN,1}$, $Layer_{FCN,2}$, $Layer_{FCN,3}$:

$$M_S, M_D, M_N = Layer_{FCN,1}(FS), Layer_{FCN,2}(FD), Layer_{FCN,3}(FN). \quad (3)$$

$M_S$, $M_D$, $M_N$ denote the processed features vector of protein sequence, description and network respectively, where $M_S, M_D, M_N$ are all $n \times h$ matrices. Then we concatenate them together and denote it as $M_{SDN}$, which is $n \times 3h$.

$$M_{SDN} = concat(M_S, M_D, M_N). \quad (4)$$

### 2.4 Protein function prediction based on GO embeddings and protein embeddings

Our model predicts the protein function by projecting GO terms and proteins into the same low-dimensional space. Let $FT \in \mathbb{R}^{z \times d_{Bert}}$ be the representation vectors of the protein function text data, where $z$ represents the number of protein functions. Let $B$ be the binary label matrix, which is $n$ by $z$. $B_{i,j} = 1$ only if protein $i$ has the function $j$. The binary cross-entropy loss function is defined as:

$$L = \sum_{i=1}^{n} \sum_{j=1}^{z} [-B_{i,j} \times log(1/1 + exp(M_{SDN}W(FT)^T)))$$
$$-(1 - B_{i,j}) \times log(exp(M_{SDN}W(FT)^T)/(1 + exp(M_{SDN}W(FT)^T)], \quad (5)$$

where $W \in R^{3h \times d_{Bert}}$ are learnable parameters. During the prediction process, we can annotate a new protein using the following equation:

$$p_j = 1/(1 + exp(m_{SDN}W(FT_j)^T)), \quad (6)$$

where we use $p_j$ to represent the probability of the new protein has the function $j$ and $m_{SDN}$ denotes this protein's concatenated features extracted by the method in **2.3**. Since our method utilizes the textual description of a new function to classify proteins, it is able to annotate a new function even if it is not annotated to any protein in the training data.

### 2.5 Annotate novel functions, sparse functions and gene sets to pathways

As for the terms seen in the training, we could combine the similarity based prediction method using DiamondScore [14,36] to enhance ProTranslator, which is denoted as ProTranslator+DiamondScore. The previous prediction score $S^j{}_{ProTranslator}$ for function $j$ is the output of the deep learning model. Therefore we redefine the overall prediction score of ProTranslator+DiamondScore on function $j$ as:

$$S^j{}_{ProTranslator+DiamondScore} = \alpha \times S^j{}_{DiamondScore} + (1 - \alpha) \times S^j{}_{ProTranslator}. \quad (7)$$

The DiamondScore is calculated as:

$$S^j{}_{DiamondScore} = \frac{\sum_{x^s \in E} bitscore(x^s{}_q, x^s) \times I(j \in J_{x^s})}{\sum_{x^s \in E} bitscore(x^s{}_q, x^s)}, \quad (8)$$

where $x^s{}_q$ is the query sequence and $E$ is the similar sequences set. $J_{x^s{}_i}$ is the annotations set of proteins with sequence feature $x^s$. $I$ is the identity function and $bitscore$ is the sequence similarity score predicted by BLAST [10].

### 2.6 Text generation by the protein sequence features

We develop a model to generate the description of proteins from the sequence features based on the Transformer architecture [37]. For each GO function, we average all the one hot encodings of the sequences of its samples as the input in the text generation model. We still leverage the convolutional kernels in DeepGOCNN [14] to process the sequence features. Then we add the main Transformer architecture. Since the DeepGOCNN model discards the positional information when setting the max pooling layer to the maximum size, we remove the positional encodings at the encoder stack bottoms in Transformer. The multi-head self-attention could be written as:

$$MultiHead(M_{SDN,l}, M_{SDN,l}, M_{SDN,l}) = concat(head_1, \ldots, head_o),$$

$$head_i = softmax(\frac{M_{SDN,l}A_i^Q(M_{SDN,l}A_i^K)^T}{\sqrt{d_k}})M_{SDN,l}A_i^V, \tag{9}$$

where $M_{SDN,l}$ represents the input of the $l_{th}$ sub-layer. $A_i^Q, A_i^K, A_i^V$ are learnable parameters. To make the optimization process more stable, we adopt the pre-layer normalization in the Transformer [38]:

$$M_{SDN,l+1} = M_{SDN,l} + \mathcal{F}(LN(M_{SDN,l}\,;\,\theta_l)), \tag{10}$$

where $SDN_l$ and $SDN_{l+1}$ represent the $l_{th}$ sub-layer's input and output. $LN$ denotes the layer normalization and $\theta_l$ represents the parameters of the sub-layer $\mathcal{F}$ in the encoder or decoder.

## 3 Experimental setup

### 3.1 Calculating similarities between GO functions

We calculated three kinds of similarities between two GO functions: text-based similarity, GO graph-based similarity and annotation-based similarity. The text-based similarity is calculated using the cosine similarity between the representation vectors of their textual descriptions. We calculated the GO graph-based similarity using the shortest distance between two GO functions on the GO graph, which is built based on "is_a" and "part_of" relationships. We calculated the annotation-based similarity by using the cosine similarity between the binary annotation vectors of two GO functions. The binary annotation vector $Annt^j \subseteq \mathbb{R}^z$ of a GO function $j$ is defined as $Annt_i^j = 1$ if function $j$ is $i$ or one of $i's$ ancestor in the GO hierarchy otherwise $Annt_i^j = 0$.

### 3.2 Datasets and evaluation

We used the Gene Ontology (GO) that was released on June 16, 2021. The descriptions in the 'def' field were used as the textual description. We considered three datasets: CAFA3 [2], SwissProt [30], and GOA(Human) [28]. The preprocessed CAFA3 challenge dataset and SwissProt dataset were obtained from the online data files provided by DeepGOPlus. The pre-trained gene network features for humans were downloaded from STRING database v9.1 [39]. We collected the gene descriptions from GeneCards [34,35]. The CAFA3 dataset was released in September, 2016. We selected the proteins from the intersection of the CAFA3 dataset, the gene network features file and the gene description file and 11,679 proteins were finally selected. The SwissProt dataset was published in January, 2016 and we finally selected 5,889 proteins. The annotations were propagated according to the hierarchical structure of GO based on "is_a" and "part_of" relationships. We collected the annotations of the GOA(Human) dataset from the Gene Ontology Consortium website. The annotation file was generated on May 1, 2021. We leveraged 3-fold cross-validation to evaluate these datasets and selected 10% of the leaf nodes in the GO graph as the novel functions in the zero-shot setting and excluded their annotations in the training dataset. We investigated the performance of ProTranslator and current state-of-art methods on annotating sparse functions with proteins less than 20 using the same GOA(Human) dataset and additional GOA(Mouse) dataset. The 3-fold cross-validation was adopted. We calculated the area under the receiver of characteristic curve (AUROC) [40] of our model on the novel and sparse functions. In the text generation, we used the bilingual evaluation understudy (BLEU) [41] score as the metric. The BLEU score was computed first between segments of generated texts and references and then averaged over them.

To classify genes into pathways, we collected the Reactome [31] and KEGG [32] pathways description and gene sets. We finally obtained 2,007 and 264 pathways in Reactome and KEGG, the average gene number in each pathway is 4.6 and 22.6 respectively. We also collected the Molecular Signatures Database (MSigDB) [33] for pathway prediction. The text in "DESCRIPTION_FULL" was selected as the text data. We evaluated our approach on pathway C2, which has the most complete textual description. There were 3,704 pathways and the average number of genes for each pathway is 3.7 in pathway C2. In each pathway, the genes in both the genesets and STRING database for humans were selected for evaluation. In the text generation part, we leveraged the GOA(Human) datasets. We select 70% functions in GO data as the training functions and 30% function as validation functions.

The input length $L$ of a protein was set to 2000 in annotating functions. We set the range of convolutional kernel size $R_D$ and $R_U$ to 8 and 128, and the step $T$ was 8. Then we could get $e = 16$ different sizes of kernels. We set $d_0$ to 512 and therefore $d_{seq}$ was 8192. The dimension of PubMedBert representations $d_{Bert}$ is 768. The dimension of Mashup representations $d_{Mashup} = 800$ and $d_{Mashup} = 1000$ for GOA(Human) and GOA(Mouse) datasets. The dimension h was set to 1500. When combining the similarity based prediction method, ProTranslator used the setting of DeepGOPlus, and $\alpha = 0.68, \alpha = 0.63, \alpha = 0.46$ for BP, MF and CC. In the Transformer architecture, we used 6 encoder layers and 6 decoder layers. We set the hidden dimension of the Transformer to 512. The attention layer heads number $o$ was 8 and $d_k$ was 64. We also used the warm-up stage during the training process. Here we set the warm-up steps to 2000. We used the greedy decode strategy in the inference process.

### 3.3 Comparison approaches

We compared our method to two comparison approaches that have the same model architecture while replacing the combined features with sequence features only. In one comparison approach, we replaced the PubMedBert embedded text vectors with Term Frequency–Inverse Document Frequency (TF-IDF) embeddings to investigate the influence of text embedding methods. In the other comparison approach, the text vectors were replaced with the ontology network vectors to make comparison with the representations of topological features. We named these two comparison approaches 'tf-idf' and 'Graph-based'. To investigate how our method performed compared with current state-of-art methods when annotating sparsely annotated functions, we selected DeepGOPlus as the comparison approaches since it has been shown to exceed multiple benchmarks in the previous research [14]. DeepGOPlus used the latest released 'alphas' when considering similarity based predictions, which are 0.68, 0.63 and 0.46 for BP, MF and CC.

## 4 Results

### 4.1 Gene Ontology term description similarity reflects function annotation similarity

The key idea of our method is to annotate a novel function by transferring annotations from other functions that have similar textual description. Therefore, we first examined the correlation between the text-based GO term similarity and the annotation-based GO term similarity (see **Experimental Setup**). We observed a strong correlation between these two similarities, indicating that GO terms with similar textual description tend to have similar protein annotations (**Fig. 1c**). We next compared the text-based similarity with GO graph-based similarity. We found that terms that are close on the graph have much higher textual similarity (**Fig. 1b**). Since previous work has demonstrated how GO graph can be used to assist function annotation, especially for sparsely annotated functions [18,42,43], the strong consistency between text-based GO similarity and GO graph-based similarity further raises our confidence that textual description can be used to enhance protein function prediction.

### 4.2 ProTranslator enables protein function prediction in the zero-shot setting

We next sought to examine whether ProTranslator can classify proteins in the zero-shot setting where the test function does not have any annotated proteins in the training data. We summarized the results of ProTranslator on three GO domains of biological process (BP), molecular function (MF) and cellular component (CC) across three datasets in **Fig. 2**. To simulate the zero-shot setting, we held out all protein annotations of test functions from the training data. We first compared ProTranslator to TF-IDF, which models the textual description using a frequency-based vector space model, and observed at least 13%, 11%, 14% improvements on BP, MF, CC domains on three datasets by using ProTranslator, indicating the superior performance of embedding text description using large-scale pre-trained language models. We then compared ProTranslator to a graph-based approach, which embeds the GO graph structure to annotate novel functions. ProTranslator also outperformed this graph-based approach by a large margin, which demonstrates the advantage of using text data against GO graphs to annotate novel functions. More importantly, the graph-based approach requires the function to be within the GO graph, whereas ProTranslator supports the annotation of any function only based on a short text description.

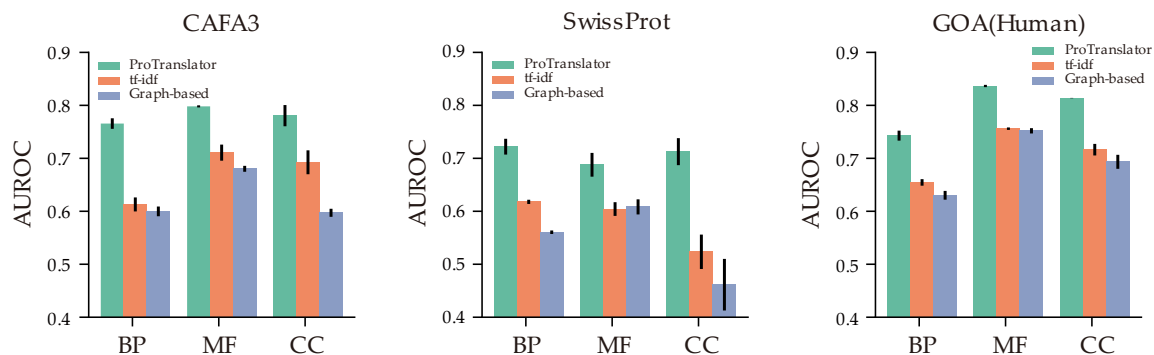

**Fig. 2 Performance of ProTranslator in the zero-shot setting.** Bar Plots comparing the AUROC of ProTranslator, tf-idf and the graph-based approach on CAFA3, SwissProt, GOA(Human). We held out all protein annotations of test functions from the training data.

### 4.3 ProTranslator obtains substantial improvement in the few-shot setting

After confirming the superior performance of ProTranslator in the zero-shot setting, we next investigated whether

ProTranslator can achieve better performance in the few-shot setting where each test GO term only has very few annotated proteins. We compared ProTranslator to the state-of-the-art approach DeepGOPlus [14] on GOA Human and Mouse datasets (**Fig. 3**). Similar to DeepGOPlus, we also incorporated the protein sequence feature into our model to improve the performance and denoted it as ProTranslator+DiamondScore. We observed substantial improvement of ProTranslator over DeepGOPlus when each GO term only has a limited number of annotation proteins between 1 to 20. We further found that the improvement of our method is larger for terms that have fewer annotated proteins, indicating the more prominent performance of our method in annotating new functions.

To further understand the effect of sequence-based features, we excluded the sequence-based DiamondScore from both our method and DeepGOPlus. DeepGOCNN [14] is the implementation of DeepGOPlus without using sequence-based DiamondScore. We observed a decreased performance for both our method and DeepGOPlus. Nevertheless, ProTranslator still outperforms DeepGOCNN with a large margin. Moreover, ProTranslator without DiamondScore also outperforms DeepGOPlus on functions that have less than 10 annotated proteins on GOA(Human), again indicating the prominent performance of using textual description for protein function prediction.

### 4.4 ProTranslator annotated genes to pathways by only using the pathway description

After observing the superior performance of ProTranslator on functions collected in the Gene Ontology, we next evaluate ProTranslator on a more challenging setting of classifying genes into a pathway without knowing any genes in that pathway. Since ProTranslator could annotate any GO term as long as its textual description is available, we hypothesized that ProTranslator could also predict genes of a given pathway by only using the description of that pathway. Specifically, we trained ProTranslator using the function annotation and text description of Gene Ontology and then applied this model to pathways in Reactome, KEGG and MSigDB. Notably, even though graph-based approaches, such as clusDCA, are able to annotate GO terms that do not have any proteins, they cannot be applied to these pathways as they require the functions to be within the GO graph. In contrast, our method does not have this restriction, as it only relies on a short description of pathways. We summarized the performance of ProTranslator on Reactome, KEGG and MSigDB pathway in **Fig. 4a-c**. To avoid potential data leakage between pathways and GO terms, we excluded pathways that have more than 90% shared genes with an existing GO term. We examined the performance of our method at different AUROC thresholds and found that our method could annotate 81% of Reactome pathways with AUROC larger than 0.85 and 84% of KEGG pathways with AUROC larger than 0.75, demonstrating the accuracy of annotating genes to pathways and functions that are not collected in the GO.

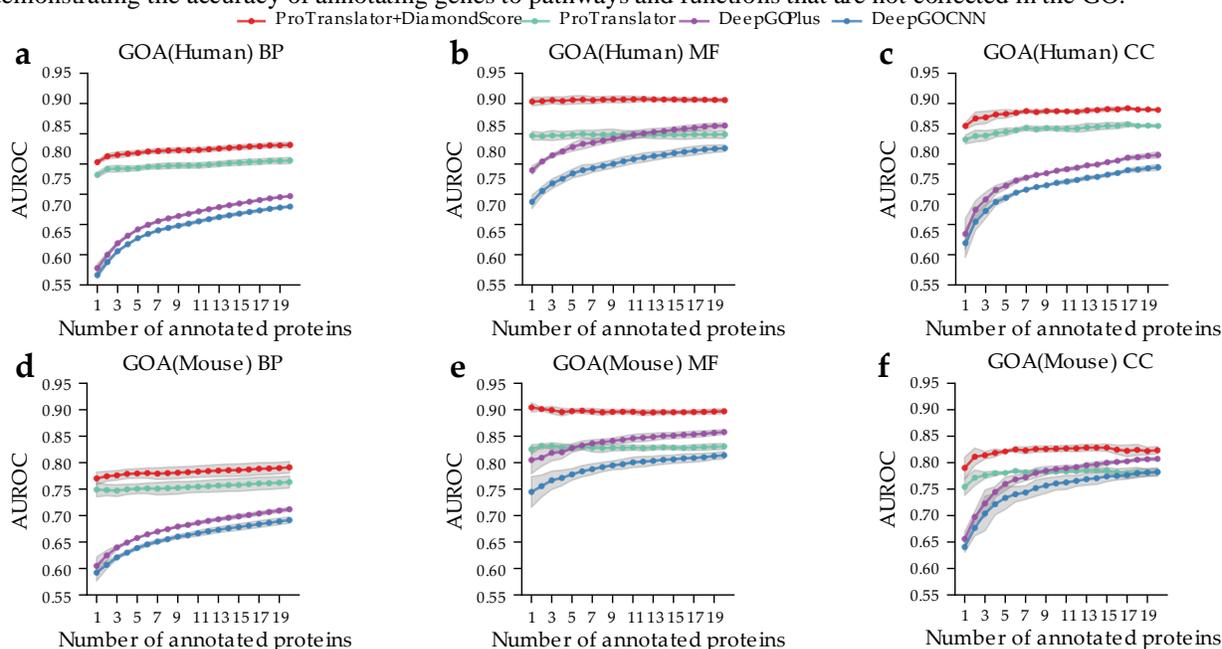

**Fig. 3. Performance of ProTranslator on sparsely annotated functions.** Plots comparing the AUROC of using ProTrainslator with sequence feature (ProTranslator+DiamondScore), ProTranslator, DeepGOPlus and DeepGOCNN on annotating functions with number of annotated proteins from 1 to 20 in the training data in GOA(Human) (**a-c**) and GOA(Mouse) (**d-f**).

To better understand the prominent performance of ProTranslator on classifying genes into these functions, we further investigated how the text information determined the performance of ProTranslator. We observed that the pathway descriptions of Reactome and KEGG dataset were closer to the GO descriptions than those of MSigDB in the embedding space (**Fig. 5a**), which explains the more prominent performance of ProTranslator on Reactome and KEGG than MSigDB. Then we plotted the three datasets separately and colored each pathway using its AUROC during the cross-validation (**Fig. 5b-d**). We observed a clear pattern that the pathway whose textual description is closer to GO description tends to have a higher AUROC. This again indicates the substantial contribution of textual description in classifying genes into pathways and shows that the performance of our method depends on the quality of the textual description.

### 4.5 ProTranslator generates text description for a gene set

The superior performance of ProTranslator comes from its novel setting of modeling the protein function prediction as a machine translation problem. We have extensively validated how to find associated proteins for a given function. Here, we aim to explore whether we can also generate the functional textual description for a set of proteins. For a given set of proteins, we used the average feature representations of them as input and then generated a novel textual description using ProTranslator. We evaluated this method on the GOA (Human) dataset by comparing the generated textual description to the ground truth curated GO term description and obtained a 0.26 BLEU. To avoid potential data leakage, we excluded the test term that has more than 0.5 Jaccard annotation-based similarity with any training GO term. By further examining the generated text, we observed that many of them are highly consistent with the curated GO term description (**Table 1**), suggesting the possibility of using our method to automatically expand GO and curate new functions.

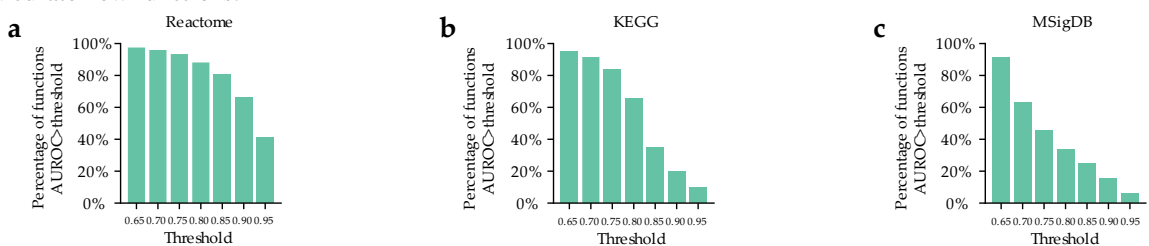

**Fig. 4. Classifying genes into pathways only based on pathway description. a-c**, Bar plots showing the percentage of pathways with AUROC greater than different thresholds (x-axis) on Reactome (a) and KEGG (b) and MSigDB (c). We didn't see any genes of these pathways in the training stage. Pathways that have many overlapped genes with an existing GO term are excluded.

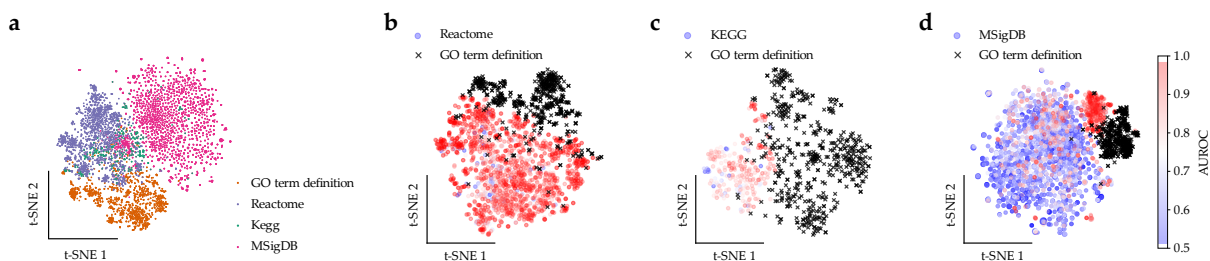

**Fig. 5. Visualization of the joint embedding space of pathways and GO terms based on textual description. a**, t-SNE plot showing the embedding space of GO terms and pathways in Reactome, KEGG, and MSigDB dataset. 1500 randomly selected GO terms are shown here. GO terms and pathways are embedded using their textual descriptions. **b,c,d,** t-SNE plots show the co-embedding space of Reactome (b), KEGG (c), MSigDB (d) and GO term. Each pathway is collected by its AUROC during the cross-validation. 500 randomly selected GO terms are shown in each plot.

**Table 1. GO term description generated by our method according to the annotated proteins. The nearest text refers to the text of the training GO term that is closest to the test GO term.**

| | GO:0032588 |
|---|---|
| **Generated text** | the lipid bilayer surrounding a vesicle transporting substances between the trans - golgi network and other parts of the cell . |
| **Nearest text in the training** | The network of interconnected tubular and cisternal structures located within the Golgi apparatus on the side distal to the endoplasmic reticulum, from which secretory vesicles emerge. The trans-Golgi network is important in the later stages of protein secretion where it is thought to play a key role in the sorting and targeting of secreted proteins to the correct destination. |
| **Ground truth text** | the lipid bilayer surrounding any of the compartments that make up the trans - golgi network . |
| | GO:0048738 |
| **Generated text** | the process whose specific outcome is the progression of a cardiac cell over time , from its formation to the mature state . a cardiac cell is a cell that will form part of the cardiac organ of an individual . |
| **Nearest text in the training** | The process in which a relatively unspecialized cell acquires the specialized structural and/or functional features of a cell that will form part of the cardiac organ of an individual. |
| **Ground truth text** | the process whose specific outcome is the progression of cardiac muscle over time , from its formation to the mature structure . |

### 4.6 Ablation experiment

ProTranslator integrated protein sequence, protein description and protein-protein network as features to embed proteins. To understand the contribution of each kind of feature, we conducted ablation studies on the GOA(Human) dataset to explore how the performances would change with different feature combinations (**Fig. 6**). We found that each of these three components makes an important contribution for the function annotation. ProTranslator could achieve the best performance with all the three features together and the AUROC was lowest with only the sequence feature. This observation verified that ProTranslator could be applied to cases where the protein sequence, description and network data were not all available.

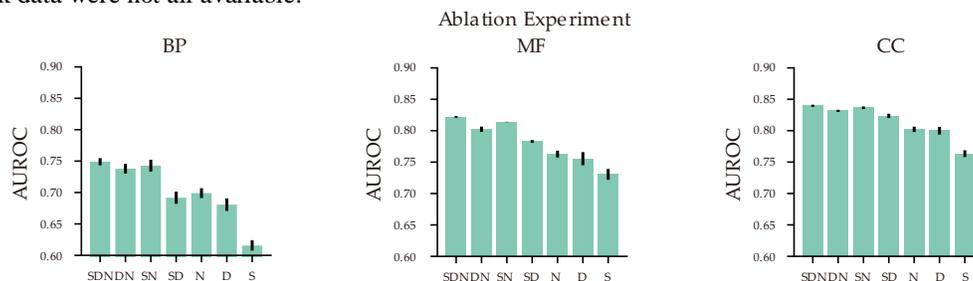

**Fig. 6. Ablations studies on contributions of different features.** Bar plots comparing the AUROC of using different feature combinations. S, D and N denote the protein sequence, description and network respectively. The tick label under each bar implies the combination, e.g., SN means using protein sequence and network features.

## 5 Conclusion and discussion

We have presented ProTranslator, a text based protein function prediction framework. Through experiments on predicting GO terms in zero-shot and few-shot settings, we have verified that our approach was able to annotate novel functions by only using textual descriptions. We have further successfully applied ProTranslator to annotate genes to pathways from Reactome, KEGG and MsigDB [31-33] using only the pathway description. We observed that the performance of ProTranslator was better for functions that have text descriptions similar to GO term descriptions. Finally, we have demonstrated how our method can be used to generate novel textual descriptions for a given set of genes, offering the possibility to automatically curate new GO terms.

Despite the novelty and prominent performance of our method, there are still a few limitations of our method. Firstly, textual description is required to annotate the new function, which could be difficult to get for an under-studied new function. We plan to incorporate genomics of these functions into our framework to supplement the text information. We will also provide interactive interfaces for users to modify their text based on the annotations provided by our method. Secondly, the AUROCs of these novel functions are relatively lower compared to AUROCs of functions that have many annotations. Predicting for densely annotated functions is known to be less challenging [18]. Although AUROC values are not very high, our method can be used to narrow down the candidate proteins for a given new function, thus substantially reducing the experimental and other validation efforts.

This work is inspired by the decade-long attempts to automatically curate Gene Ontology, including NeXO [44] and CliXO [45]. The key difference between us and these pioneering works is that they reconstructed the hierarchical structure and gene clusters in the GO, whereas we generate the textual description of terms. These textual descriptions play a key role in scientific communication and collaborations and their curation is often most labor-intensive. Our method complements these existing efforts by using a novel natural language processing perspective and fills in an important gap towards automating GO curation. Another line of related works is automatically generating the term name for a set of genes or proteins [46-48]. Compared to these approaches, we generate a free text that contains a few sentences, which are more informative than a simple term name. Moreover, these existing approaches restricted the generated term to be a known phrase in the existing literature, whereas the text generated by our method is de novo, thus offering the unique description to a novel function or pathway.